\documentclass[a4paper]{aa}

\usepackage{graphicx,natbib,amsmath}
\usepackage{txfonts}
\usepackage{color}
\usepackage[usenames,dvipsnames]{xcolor}

\newcommand{\bz}{$\langle B_\mathrm{z} \rangle$}

\newcommand{\vtau}{V410~Tau}
\newcommand{\hd}{HD\,37776}

\newcommand{\kms}{km\,s$^{-1}$}

\newcommand{\firps}[1]{\resizebox{\hsize}{!}{\rotatebox{90}{\includegraphics{#1}}}}
\newcommand{\fifps}[2]{\centering\resizebox{#1}{!}{\includegraphics{#2}}}

\newcommand{\finps}[3]{\centering\resizebox{#1}{#2}{\includegraphics{#3}}}

\newcommand{\beq}{\begin{equation}}
\newcommand{\eeq}{\end{equation}}

\begin{document}

\title{Diagnostic of stellar magnetic fields with cumulative \\ circular polarisation profiles%
%\thanks{Based on observations obtained at the Canada-France-Hawaii Telescope (CFHT) which is operated by the National Research Council of Canada, the Institut National des Sciences de l'Univers of the Centre National de la Recherche Scientifique of France, and the University of Hawaii.}
}

\author{O.~Kochukhov%\inst{1}
}

\institute{
Department of Physics and Astronomy, Uppsala University, Box 516, 75120 Uppsala, Sweden
}

\date{Received 15 April 2015 / Accepted 22 May 2015}

\titlerunning{Cumulative Stokes $V$ profiles}
\authorrunning{O. Kochukhov}

\abstract%
{
Information about stellar magnetic field topologies is obtained primarily from high-resolution circular polarisation (Stokes $V$) observations. Due to their generally complex morphologies, the stellar Stokes $V$ profiles are usually interpreted with elaborate inversion techniques such as Zeeman Doppler imaging (ZDI). Here we further develop a new method of interpretation of circular polarisation signatures in spectral lines using cumulative Stokes $V$ profiles (anti-derivative of Stokes $V$). This method is complimentary to ZDI and can be applied for validation of the inversion results or when the available observational data are insufficient for an inversion. Based on the rigorous treatment of polarised line formation in the weak-field regime, we show that, for rapidly rotating stars, the cumulative Stokes $V$ profiles contain information about the spatially resolved longitudinal magnetic field density. Rotational modulation of these profiles can be employed for a simple, qualitative characterisation of the stellar magnetic field topologies. We apply this diagnostic method to the archival observations of the weak-line T~Tauri star \vtau\ and Bp He-strong star \hd. We show that the magnetic field in \vtau\ is dominated by an azimuthal component, in agreement with the ZDI map that we recover from the same data set. For \hd\ the cumulative Stokes $V$ profile variation indicates the presence of multiple regions of positive and negative field polarity. This behaviour agrees with the ZDI results but contradicts the popular hypothesis that the magnetic field of this star is dominated by an axisymmetric quadrupolar component.
}


\maketitle

\section{Introduction}
\label{intro}

Stellar magnetism represents an important, though not fully understood and poorly constrained, ingredient of the theories of stellar formation and evolution. Although magnetic fields are believed to play an important role in many astrophysical situations, their direct detection and characterisation is often very challenging. The stellar surface magnetic fields with typical strengths ranging from a few Gauss to several kilo-Gauss are commonly diagnosed with the help of the Zeeman effect. The broadening and splitting of magnetically sensitive spectral lines lines becomes apparent in high-resolution spectra when the field strength exceeds $\sim$\,1~kG, allowing one to measure the mean field modulus \citep{mathys:1997b,kochukhov:2006b,reiners:2007} and identify different field components \citep{johns-krull:1996,shulyak:2014} in strongly magnetised objects. However, this Zeeman broadening analysis is limited to slow rotators and provides little information about the magnetic field geometry. On the other hand, analysis of circular (and more recently linear) polarisation in spectral lines enables detection of much weaker magnetic fields \citep{wade:2000b,petit:2011,marsden:2014}, especially when polarisation signals can be enhanced with line-addition methods \citep{donati:1997,kochukhov:2010a}. Reconstruction of the detailed surface magnetic field vector maps with the Zeeman Doppler imaging (ZDI) technique heavily relies on the interpretation of polarisation in spectral line profiles \citep{brown:1991,piskunov:2002a,carroll:2012}.

For early-type magnetic stars, which typically host strong, globally organised, dipolar-like fossil fields, one frequently observes morphologically simple, e.g. two-lobe, S-shaped, circular polarisation (Stokes $V$) signatures. Several integral magnetic observables, for example the mean longitudinal magnetic field \citep{mathys:1991} and crossover \citep{mathys:1995a}, can be derived by computing the wavelength moments of such simple Stokes $V$ profiles. These observables are interpreted by fitting their rotational phase curves with low-order multipolar field models \citep{landstreet:2000,bagnulo:2002}. Although this technique misses some small-scale surface magnetic field structures \citep{kochukhov:2004d,kochukhov:2010}, it is straightforward to apply, computationally inexpensive and therefore is widely used for obtaining information on the global magnetic field geometries of large stellar samples \citep[e.g.][]{auriere:2007,hubrig:2007b}.

In contrast to stable and usually topologically simple magnetic fields of early-type stars, the dynamo-degenerated surface fields of active late-type stars are weak, complex and rapidly evolving. Their circular polarisation profile shapes are correspondingly more complex, often exhibiting many lobes. Such Stokes $V$ profiles cannot be meaningfully characterised with integral magnetic observables because their low-order moments tend to be close to zero even when strong polarisation signatures are evident in the data \citep[e.g.][]{kochukhov:2013}. In other words, the surface magnetic field distribution is so complex that vector averaging over the visible stellar disk leads to a substantial cancellation of the Stokes $V$ signal corresponding to different field polarities. In this situation one usually resorts to an elaborate ZDI modelling of the Stokes $V$ profiles themselves, which imposes stringent requirements on the signal-to-noise ratio and rotational phase coverage of the observational data. Being an intrinsically ill-posed inversion problem, reconstruction of the stellar magnetic field topologies with ZDI suffers from a number of uniqueness and reliability issues \citep{donati:1997a,kochukhov:2002c,rosen:2012}, not the least related to the fact that only circular polarisation rather than the full Stokes vector spectra are typically available for active late-type stars. In this respect, ZDI is often perceived to be less reliable in comparison to the usual mapping of temperature spots or chemical inhomogeneities using intensity spectra. Due to a complex response of the circular polarisation spectra to the strength and orientation of the local magnetic field one cannot establish a straightforward connection between the polarisation profile variability pattern and major surface magnetic features in the same way as, for example, one can recognise individual star spots in the dynamic intensity spectra \citep[e.g.][]{barnes:2000}. It is therefore of great interest to devise simple, yet informative, alternative methods of the analysis of complex Stokes $V$ profiles that can be used to validate ZDI results or applied when the rotational phase coverage is insufficient for a magnetic inversion.

Several recent studies proposed new methods of extracting information from the circular polarisation signatures. \citet{carroll:2014} considered diagnostic potential of the net absolute Stokes $V$ signal, showing that under certain assumptions it allows to characterise the apparent absolute longitudinal magnetic field. On the other hand, \citet{gayley:2014} suggested to use an anti-derivative of the Stokes $V$ profile for a more intuitive interpretation of the circular polarisation data. In this paper we further develop the latter idea. The main goal of this work is to present a comprehensive theoretical formulation of the new polarisation observable, demonstrate its connection to the underlying stellar surface magnetic field structure and to apply the new magnetic field diagnostic method both to simulated circular polarisation data and to real spectropolarimetric observations of stars with topologically complex magnetic fields.

\section{Cumulative Stokes $V$ profiles}

\begin{figure*}[!th]
\centering
\finps{!}{8.5cm}{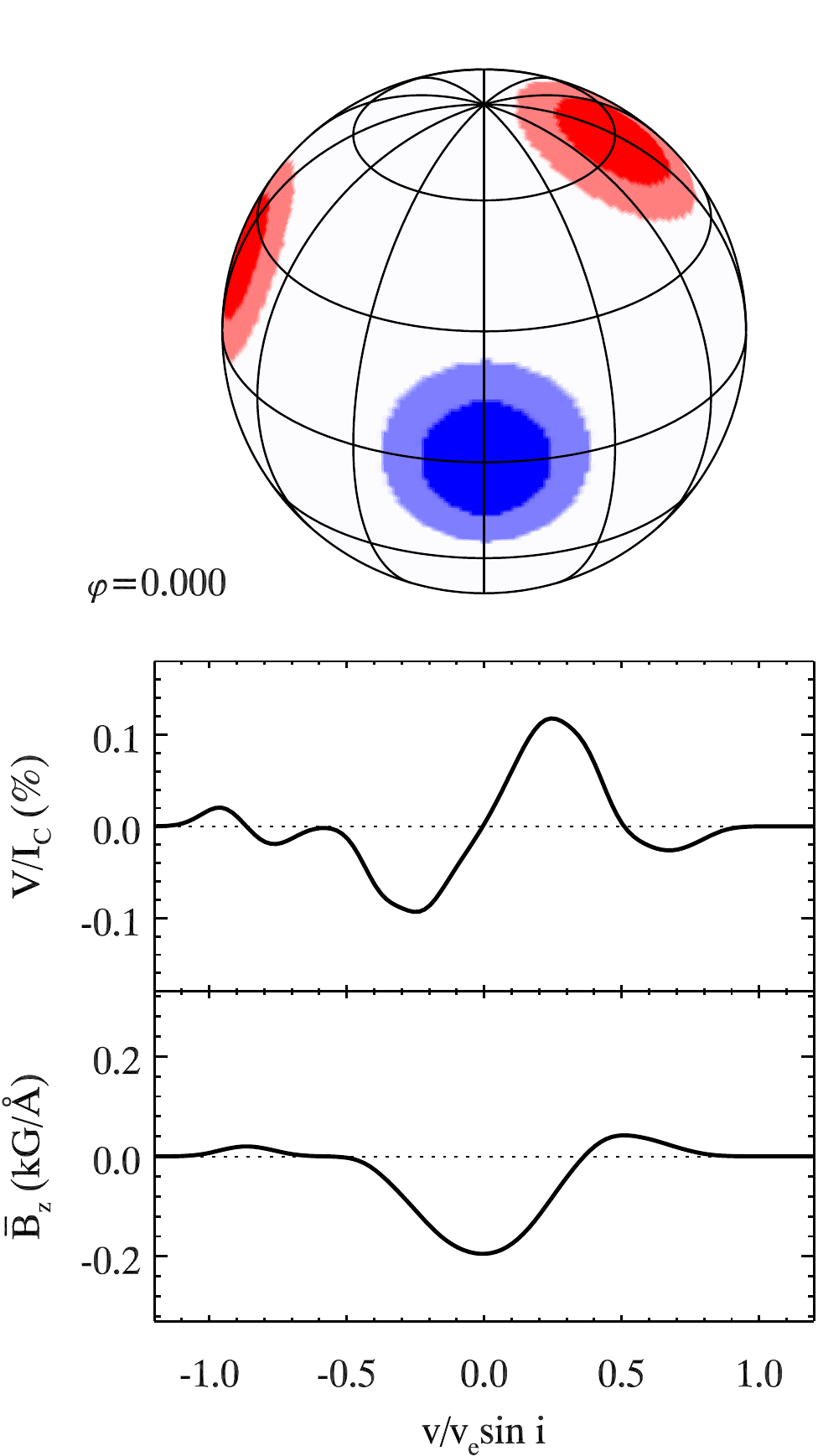}\hspace*{0.5cm}
\finps{!}{8.5cm}{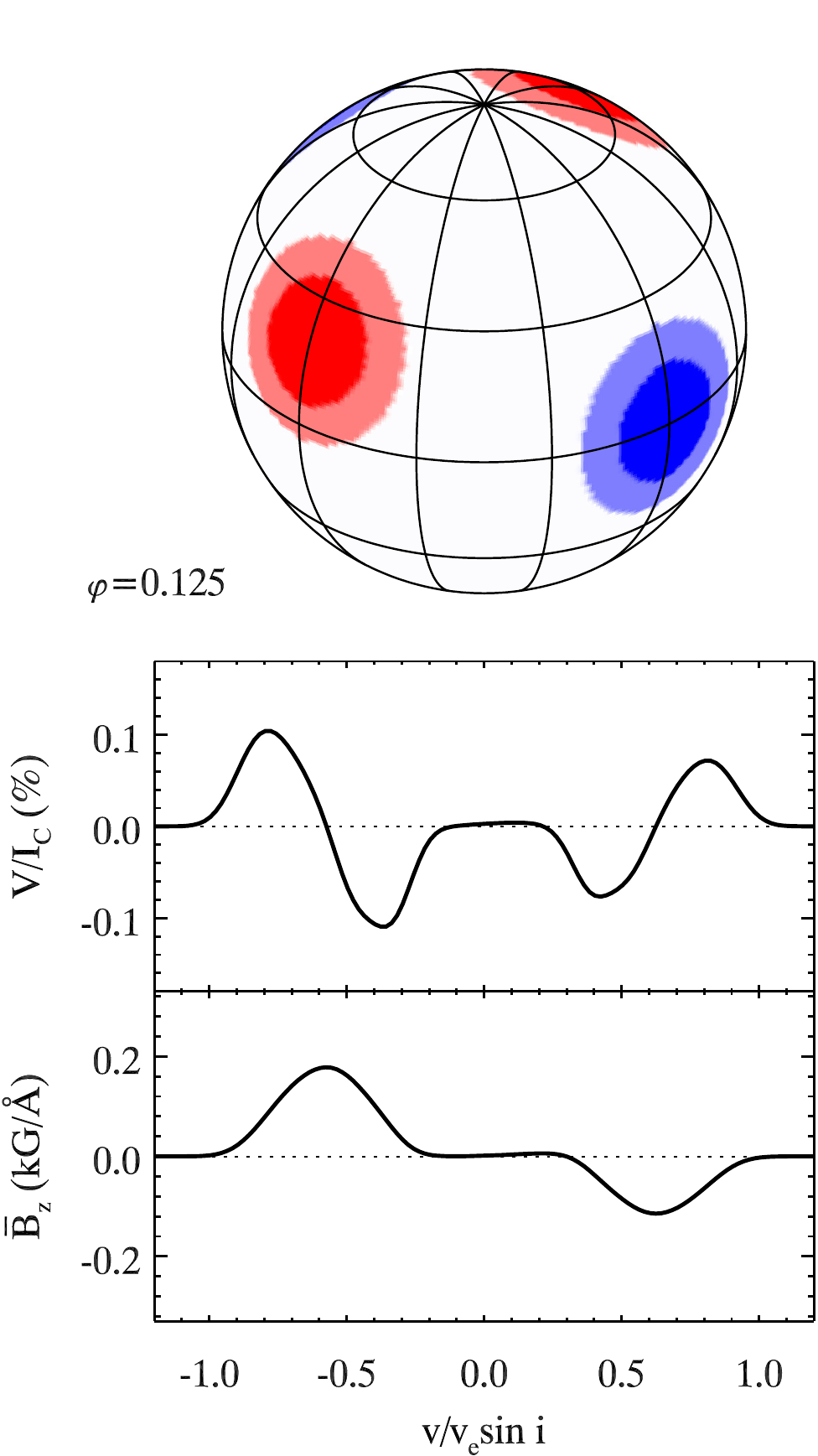}\hspace*{0.5cm}
\finps{!}{8.5cm}{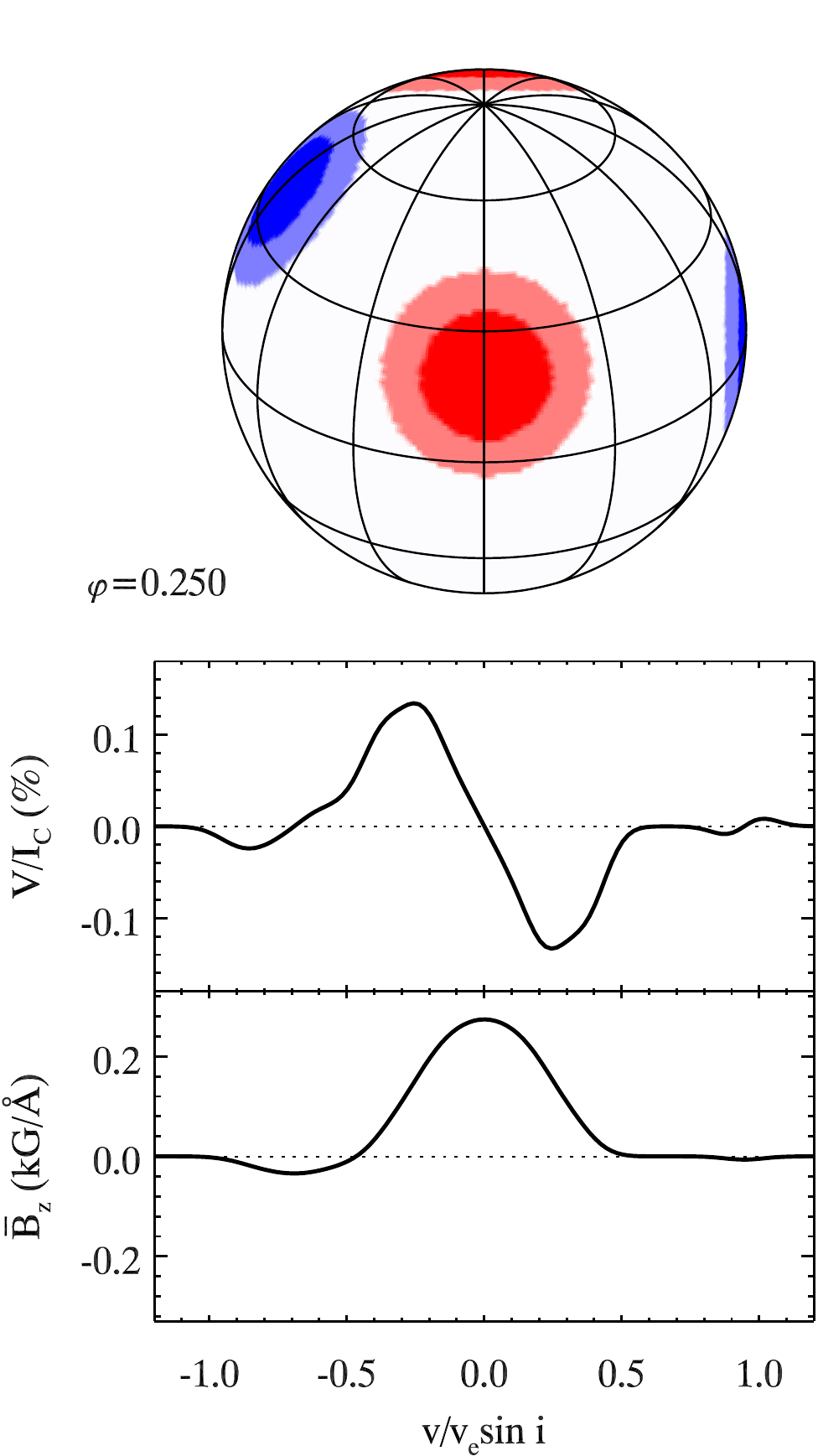}
\caption{Model Stokes $V$ (upper curves) and cumulative Stokes $V$ (lower curves) profiles for a star with several large circular magnetic spots of different polarities. The line profiles and the spherical maps of the radial magnetic field component are shown for three different rotational phases. The phases are indicated above each line profile panel.}
\label{fig:model}
\end{figure*}

\subsection{Theoretical basis}
\label{core_theory}

To describe the disk-integrated Stokes parameter profiles of a rotating star we consider a coordinate system with the z-axis directed towards the observer and the y-axis located in the plane formed by the line-of-sight and the stellar rotational axis. The star is assumed to have a unit radius and to rotate counterclockwise as seen from the visible rotational pole. Then, the Doppler shift across the stellar disk is a function of the x-coordinate alone. With these conventions the disk-integrated continuum intensity is 
\begin{equation}
{\cal F}_c = \int_{-1}^{+1} dx \int_{-\sqrt{1-x^2}}^{+\sqrt{1-x^2}} I_c(x,y) dy
\end{equation}
and the disk-integrated Stokes profiles of a spectral line with the central wavelength $\lambda_0$ are represented by the integrals
\begin{equation}
{\cal F}_I = \int_{-1}^{+1} dx \int_{-\sqrt{1-x^2}}^{+\sqrt{1-x^2}} I[x,y;\lambda-\lambda_0-\Delta\lambda_R x; \boldsymbol{B}(x,y)] dy
\end{equation}
and
\begin{equation}
{\cal F}_V = \int_{-1}^{+1} dx \int_{-\sqrt{1-x^2}}^{+\sqrt{1-x^2}} V[x,y;\lambda-\lambda_0-\Delta\lambda_R x; \boldsymbol{B}(x,y)] dy,
\end{equation}
where $I_c(x,y)$ represents the local continuum intensity and the local line intensity and polarisation are given by $I(x,y,\lambda,\boldsymbol{B})$ and $V(x,y,\lambda,\boldsymbol{B})$. The amplitude of the Doppler shift due to the stellar rotation is determined by the projected rotational velocity $v_{\rm e}\sin{i}$
\begin{equation}
\Delta\lambda_R=\dfrac{\lambda_0 v_{\rm e}\sin{i}}{c}.
\end{equation}

Under the weak-field approximation \citep[e.g.][]{polarization:2004} the local intensity profile is unaffected by the magnetic field while the Stokes $V$ profile is determined by the product of the first derivative of Stokes $I$ and  the line-of-sight component $B_z(x,y)$ of the local magnetic field vector
\begin{equation}
V = -C_Z \bar{g} B_z(x,y) \dfrac{\partial I}{\partial \lambda}.
\label{eq5}
\end{equation}
In this expression $\bar{g}$ denotes the effective Land\'e factor of a spectral line and
\begin{equation}
C_Z = \dfrac{e\lambda_0^2}{4\pi m_{\mathrm e} c^2} = 4.6686\times10^{-13} \lambda_0^2
\end{equation}
for the field measured in G and wavelength in \AA.

Using this approximation of the local Stokes $V$ profile one can express the disk-integrated circular polarisation spectrum as
\begin{equation}
\begin{split}
{\cal F}_V & = -C_Z \bar{g} \displaystyle\int_{-1}^{+1} dx \int_{-\sqrt{1-x^2}}^{+\sqrt{1-x^2}} B_z(x,y)  \\
& \times \dfrac{\partial }{\partial \lambda} I[x,y;\lambda-\lambda_0-\Delta\lambda_R x] dy. 
\end{split}
\end{equation}
Recalling that the observed Stokes parameter profiles are normalised by the disk-integrated continuum intensity, we define the normalised disk-integrated spectral line intensity
\begin{equation}
{\cal R}_I\equiv\dfrac{{\cal F}_I}{{\cal F}_c} = \dfrac{\displaystyle \int_{-1}^{+1} dx \int_{-\sqrt{1-x^2}}^{+\sqrt{1-x^2}} I[x,y;\lambda-\lambda_0-\Delta\lambda_R x] dy}
{\displaystyle \int_{-1}^{+1} dx \int_{-\sqrt{1-x^2}}^{+\sqrt{1-x^2}} I_c(x,y) dy}
\end{equation}
and circular polarisation
\begin{equation}
\begin{split}
{\cal R}_V\equiv & \dfrac{{\cal F}_V}{{\cal F}_c} = -\dfrac{C_Z \bar{g}}{\displaystyle \int_{-1}^{+1} dx \int_{-\sqrt{1-x^2}}^{+\sqrt{1-x^2}} I_c(x,y) dy} \\
& \times \displaystyle \int_{-1}^{+1} dx \int_{-\sqrt{1-x^2}}^{+\sqrt{1-x^2}} B_z(x,y)  \dfrac{\partial }{\partial \lambda} I[x,y;\lambda-\lambda_0-\Delta\lambda_R x] dy.
\end{split}
\end{equation}
The latter equation can be equivalently written as
\begin{equation}
\begin{split}
{\cal R}_V = & \dfrac{\partial }{\partial \lambda} \Bigg\{  \dfrac{ C_Z \bar{g}}{{\cal F}_c} 
 \int_{-1}^{+1} dx \int_{-\sqrt{1-x^2}}^{+\sqrt{1-x^2}} B_z(x,y)  \\
 & \times \left (I_c(x,y) - I[x,y;\lambda-\lambda_0-\Delta\lambda_R x] \right) dy \Bigg\}
\end{split}
\end{equation}
or
\begin{equation}
\begin{split}
{\cal R}_V = & \dfrac{\partial }{\partial \lambda} \Bigg\{ C_Z \bar{g} 
 \int_{-1}^{+1} dx \int_{-\sqrt{1-x^2}}^{+\sqrt{1-x^2}} B_z(x,y) \\
 & \times  i_c(x,y) r_I[x,y;\lambda-\lambda_0-\Delta\lambda_R x] dy \Bigg\},
\end{split}
\label{eq11}
\end{equation}
where $i_c$ denotes the local normalised continuum intensity
\begin{equation}
i_c (x,y) \equiv  \dfrac{I_c(x,y)}{{\cal F}_c}
\end{equation}
and $r_I$ corresponds to the local normalised residual Stokes $I$ profile
\begin{equation}
r_I [x,y;\lambda-\lambda_0-\Delta\lambda_R x] \equiv \dfrac{I_c(x,y) - I[x,y;\lambda-\lambda_0-\Delta\lambda_R x]}{I_c(x,y)}.
\end{equation}

We define \textit{the cumulative Stokes $V$ (CSV) profile} as
\begin{equation}
{\cal G}_V (\lambda) \equiv \dfrac{1}{W_I} \int_{\lambda_{\rm min}}^{\lambda} {\cal R}_V(\lambda^\prime) d \lambda^\prime,
\label{eq14}
\end{equation}
where 
\begin{equation}
W_I \equiv \int_{\lambda_{\rm min}}^{\lambda_{\rm max}} (1-{\cal R}_I) d\lambda
\end{equation}
is the equivalent width of the disk-integrated Stokes $I$ profile and the integration limits $\lambda_{\rm min}$, $\lambda_{\rm max}$ cover the full extent of a spectral line. Substituting Eq.~(\ref{eq11}) into Eq.~(\ref{eq14}) yields
\begin{equation}
\begin{split}
{\cal G}_V (\lambda) = & \dfrac{ C_Z \bar{g} }{W_I}
 \int_{-1}^{+1} dx \int_{-\sqrt{1-x^2}}^{+\sqrt{1-x^2}} B_z(x,y)  \\
 & \times i_c(x,y) r_I[x,y;\lambda-\lambda_0-\Delta\lambda_R x] dy. 
\end{split}
\end{equation}
Dividing this quantity by $C_Z \bar{g}$ provides \textit{the longitudinal magnetic field density}
\begin{equation}
\begin{split}
\overline{B}_z(\lambda) \equiv \dfrac{{\cal G}_V (\lambda)}{C_Z \bar{g}} = &
\dfrac{1}{W_I}
 \int_{-1}^{+1} dx \int_{-\sqrt{1-x^2}}^{+\sqrt{1-x^2}} B_z(x,y)  \\
& \times i_c(x,y) r_I[x,y;\lambda-\lambda_0-\Delta\lambda_R x] dy. 
\end{split}
\label{eq17}
\end{equation}
This quantity represents a velocity-resolved measure of the line-of-sight magnetic field component, weighted by the projected surface area and by the equivalent width-normalised local Stokes $I$ profile. $\overline{B}_z(\lambda)$ has the units of magnetic field divided by wavelength. Its integration over the full line profile gives the normalised first moment of Stokes $V$
\begin{equation}
\int_{\lambda_{\rm min}}^{\lambda_{\rm max}} \overline{B}_z(\lambda) d\lambda = -\dfrac{1}{C_Z \bar{g} W_I} \int_{\lambda_{\rm min}}^{\lambda_{\rm max}} (\lambda-\lambda_0) {\cal R}_V d\lambda \equiv \langle B_z \rangle,
\end{equation}
commonly known as the mean longitudinal magnetic field.

The ${\cal G}_V (\lambda)$ observable discussed above is identical to the Stokes $V$ anti-derivative concept introduced by \citet{gayley:2014} with the exception that we formulate this quantity with respect to the normalised Stokes $V$ parameter ${\cal R}_V$ and divide the resulting profiles by $W_I$ whereas the earlier study proposed to divide by $1-{\cal R}_I$. The quantity ${\cal G}^{\,\prime}_V (\lambda)$ discussed by \citet{gayley:2014} is easily recovered from our CSV profiles using the transformation
\begin{equation}
{\cal G}^{\,\prime}_V (\lambda) = \dfrac{W_I}{1-{\cal R}_I} {\cal G}_V (\lambda).
\label{eq19}
\end{equation}
The corresponding \textit{velocity-resolved mean longitudinal magnetic field}  $\overline{B}^{\,\prime}_z(\lambda)$, which can be obtained by dividing ${\cal G}^{\,\prime}_V (\lambda)$ by $C_Z \bar{g}$, has the units of magnetic field strength.

The longitudinal magnetic field density $\overline{B}_z(\lambda)$ or the velocity-resolved longitudinal field $\overline{B}^{\,\prime}_z(\lambda)$ derived from the CSV profiles represent morphologically simpler observables compared to the Stokes $V$ profiles themselves. They characterise the line-of-sight magnetic field averaged along the stripes of constant Doppler shift, thus providing a more intuitive representation of the information content of the Stokes $V$ spectra.

Although the observables $\overline{B}_z(\lambda)$ and $\overline{B}^{\,\prime}_z(\lambda)$ are closely related, different normalisation choices lead to a somewhat different behaviour. The velocity-resolved mean longitudinal field closely tracks the actual local line-of-sight magnetic field component. The longitudinal field density includes an additional weighting by the projected surface area and therefore gradually diminishes to zero towards the line profile edges even if the underlying $B_z$ distribution is uniform. The $\overline{B}^{\,\prime}_z(\lambda)$ quantity compensates changing projected area by the $1-{\cal R}_I$ normalisation but instead suffers from major noise artefacts at profile edges due to division of two small numbers. Compared to $\overline{B}_z(\lambda)$, this quantity may also be adversely affected by the distortions of Stokes $I$ profile shape caused by stellar surface inhomogeneities.

Equation~(\ref{eq17}) can be considered as a convolution integral with a kernel given by the residual Stokes $I$ profile. The larger is the ratio of $\Delta\lambda_R$ to the local profile width $\Delta\lambda_l$, the less susceptible is $\overline{B}_z(\lambda)$ to the cancellation of polarisation signatures corresponding to different field polarities. Therefore, the new magnetic observable will be most effective for rapid rotators since in this case $\Delta\lambda_R\gg\Delta\lambda_l$.

Based on these definitions, we carried out an illustrative calculation of the $\overline{B}_z(\lambda)$ profiles for a magnetic field distribution comprising several circular spots with the radial field strength of 0.5--1.0~kG (see Fig.~\ref{fig:model}). The calculations assumed a Gaussian shape with $FWHM=8$~\kms\ for the local Stokes $I$ profile. The adopted central wavelength and effective Land\'e factor, $\lambda_0=5630$~\AA\ and $\bar{g}=1.215$, corresponded to the mean values of the LSD line mask used for \vtau\ (see Sect.~\ref{obs}). The star was assumed to rotate with $v_{\rm e}\sin{i}=50$~\kms\ and to be inclined by $i=60\degr$ with respect to the line of sight.

Figure~\ref{fig:model} shows the spherical maps of the radial magnetic field as well as the Stokes $V$ and CSV profile for three different rotational phases. When the surface magnetic field distribution is dominated by one of the magnetic polarities, the Stokes $V$ profile exhibits the classical S-shape pattern. The corresponding CSV profiles show a single bump, which can be either positive or negative, depending on the dominant magnetic field polarity. On the other hand, when several regions of different polarity are present on the stellar disk (e.g. middle panel in Fig.~\ref{fig:model}), the Stokes $V$ profile exhibits multiple components, which usually cannot be interpreted in terms of the surface magnetic field distribution without carrying out a detailed modelling. In contrast, the CSV profiles for the same rotational phase are simpler and readily show spots of different polarity at the longitudes corresponding to the Doppler shift within a spectral line profile.

\begin{figure}[!th]
\centering
\fifps{\hsize}{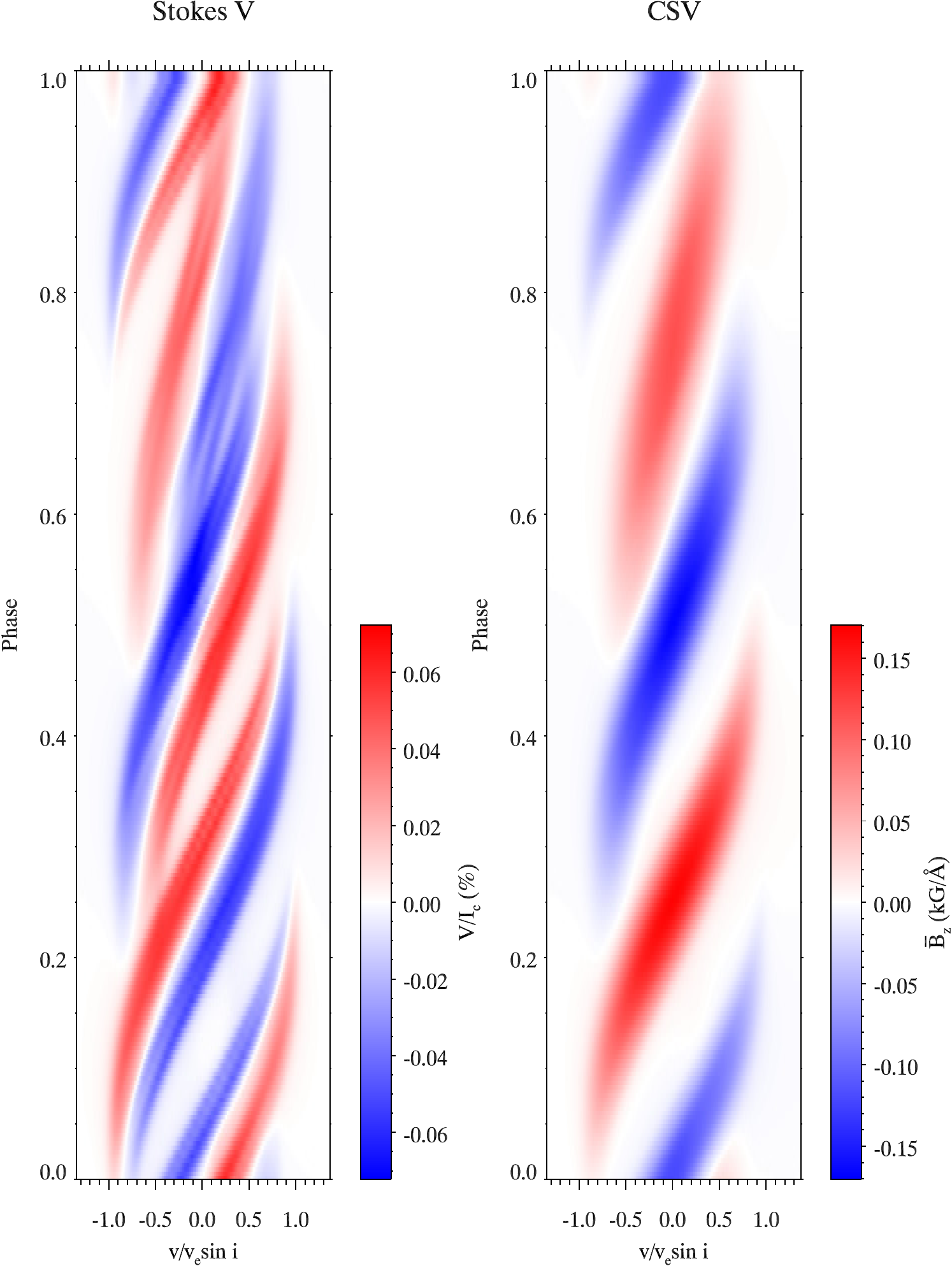}
\caption{Theoretical dynamic Stokes $V$ (left panel) and cumulative Stokes $V$ (right panel) profiles corresponding to the radial magnetic field maps shown in Fig.~\ref{fig:model}.}
\label{fig:dynamic_model}
\end{figure}

\subsection{Deconvolved longitudinal magnetic field density}

A simplified form of Eq.~(\ref{eq17}) can be obtained by assuming that the local residual intensity profile is represented by a Gaussian function which does not vary across the stellar surface. This approximation, together with the weak-field assumption, is commonly made in the context of modelling stellar circular polarisation spectra \citep[e.g.][]{petit:2004a,petit:2012}. Then, omitting $\lambda_0$, the normalised residual profile becomes
\begin{equation}
\dfrac{ r_I (\lambda,x)}{W_I} = \dfrac{1}{\sigma\sqrt{2\pi}}\exp{\left[-\frac{(\lambda-\Delta\lambda_R x)^2}{2\sigma^2}\right]}.
\end{equation}
Since the kernel described by this equation is known once $\sigma$ and $v_{\rm e}\sin{i}$ are specified, it is possible to define \textit{deconvolved longitudinal magnetic field density}
\begin{equation}
\overline{B}^{\star}_z(\lambda) = \overline{B}^{\star}_z(\Delta\lambda_R x) =  \int_{-\sqrt{1-x^2}}^{+\sqrt{1-x^2}} B_z(x,y)  i_c(x,y) dy,
\end{equation}
in which the effect of averaging over the line profile width is taken out.

In practice, $\overline{B}^{\star}_z(\lambda)$ can be obtained from the observed $\overline{B}_z(\lambda)$ with the help of a suitable deconvolution algorithm. For the interpretation of $\overline{B}^{\star}_z(\lambda)$ it can also be assumed that the centre-to-limb variation of the continuum intensity is described by a linear limb-darkening law
\begin{equation}
 i_c(x,y) = \dfrac{3(1 - \varepsilon + \varepsilon \mu)}{\pi(3-d)},
\end{equation}
where $\varepsilon$ is a limb-darkening coefficient and $\mu\equiv\sqrt{1-x^2-y^2}$.

\begin{figure}[!th]
\centering
\fifps{8cm}{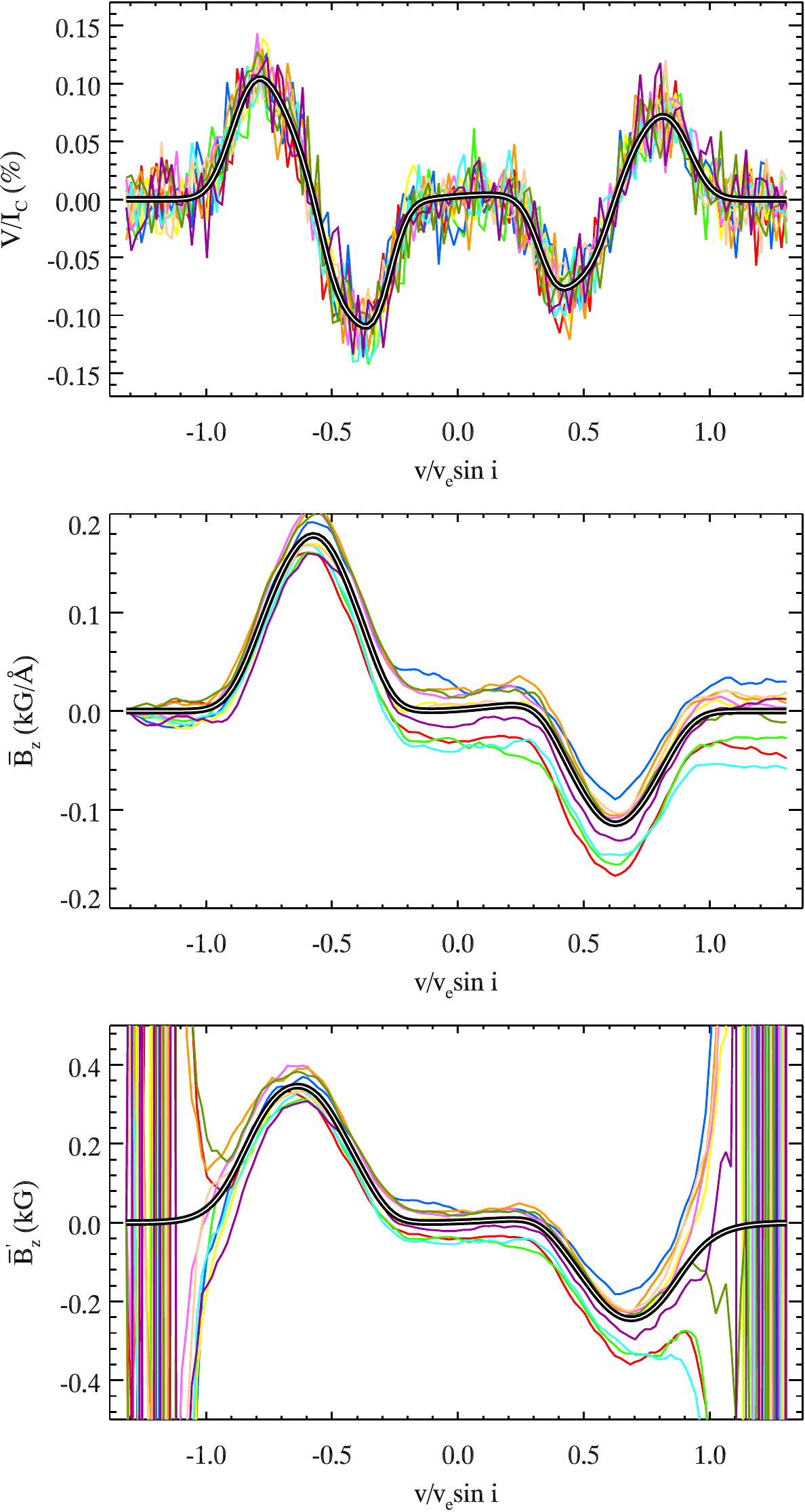}
\caption{Effect of a random noise with $\sigma=2\times10^{-4}$ on the Stokes $V$ (\textit{upper panel}) and CSV profiles with $W_I$ (\textit{middle panel}) and $1-{\cal R}_I$ (\textit{bottom panel}) normalisations. The thick double line shows the noise-free calculations corresponding to the rotational phase 0.125 of the model Stokes profiles discussed in Sect.~\ref{core_theory}. The thin lines represent profiles for different noise realisations. The $\overline{B}_z$ and $\overline{B}^{\,\prime}_z$ profiles are obtained with the forward integration according to Eq.~(\ref{eq14}).}
\label{fig:noise}
\end{figure}

\subsection{Dynamic CSV profiles}

A two-dimensional plot of the Steaks $I$ profile variability pattern as a function of wavelength (or velocity) and rotational phase is commonly used to assess distribution of the stellar surface inhomogeneities. Although similar plots of the Stokes $V$ profile variation are also occasionally published \citep{donati:1999,donati:2006}, they could not be directly employed for obtaining information about the stellar surface magnetic field distributions. The \textit{dynamic cumulative Stokes $V$ profiles}, on the other hand, allow one to trace individual magnetic spots and characterise the stellar magnetic field geometry. 

An example of the dynamic CSV profiles is shown in Fig.~\ref{fig:dynamic_model}. This figure presents polarisation spectra and corresponding $\overline{B}_z$ profiles for the model magnetic field distribution discussed in Sect.~\ref{core_theory}. The longitudinal positions and polarities of the four magnetic spots can be directly read out from the dynamic CSV plot. The latitudes of the spots can be deduced from the velocity span of their signatures in the $\overline{B}_z$ profile.

\subsection{Deriving CSV profiles from noisy data}

\begin{figure}[!th]
\centering
\fifps{8cm}{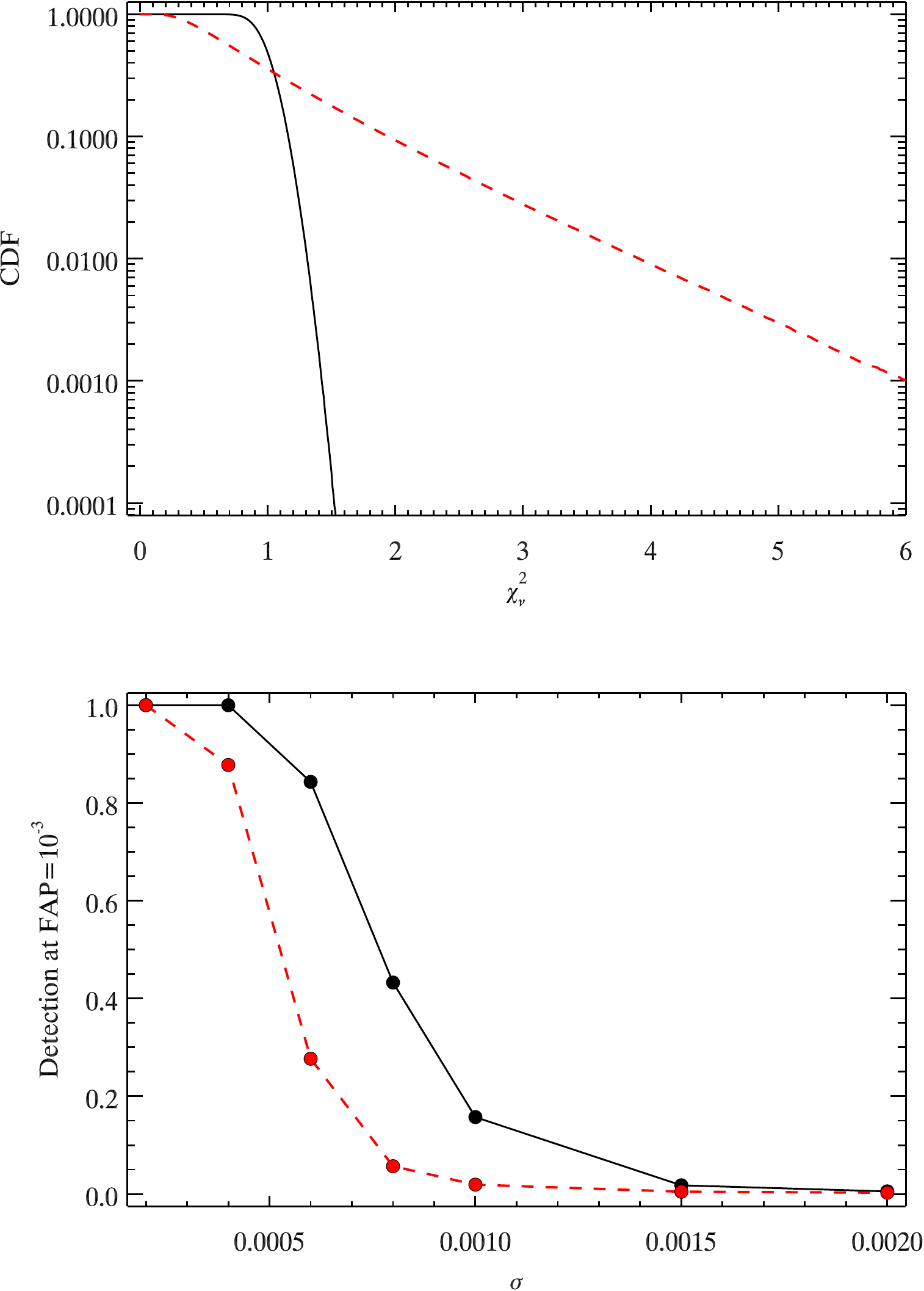}
\caption{\textit{Upper panel:} cumulative distribution of $\chi^2_\nu$ values for a set of independent, normally-distributed random variables (solid line) and for $\chi^2_\nu$ corresponding to $\overline{{\cal G}_V} (\lambda)$ given by Eq.~(\ref{eq24}) (dashed line). \textit{Lower panel:} fraction of detections of magnetic field signatures in the Stokes $V$ (solid line) and CSV profiles (dashed line) as a function of random noise added to the model Stokes profiles.}
\label{fig:chi}
\end{figure}

The cumulative Stokes $V$ profiles exhibit non-trivial noise properties. On the one hand, since the CSV spectra are computed by integrating the Stokes $V$ signatures, one may expect some cancellation of the random noise. On the other hand, the resulting noise in the CSV profiles themselves is highly correlated, leading to a qualitatively different behaviour compared to the initial Stokes $V$ profiles. These special noise properties have to be considered when applying CSV diagnostic to real observational data.

We investigated the noise properties of the CSV profiles with several sets of Monte-Carlo simulations. First, we considered the model Stokes $V$ spectra corresponding to the four-spot magnetic field distribution discussed in Sect.~\ref{core_theory}. The mean peak-to-peak amplitude of these circular polarisation profiles is $\approx$\,$2\times10^{-3}$. The profiles were sampled with a step of 1~\kms, yielding about 100 spectral points across the line. A random, normally-distributed noise with $\sigma=2\times10^{-4}$ was added to these profiles. Typical CSV signatures resulting from the forward integration according to Eq.~(\ref{eq14}) are presented in Fig.~\ref{fig:noise}. It is evident that, while the CSV spectra \textit{appear} to have less random noise compared to the initial Stokes $V$ profiles, they suffer from a systematic, ramping deviation from the zero line, which increases as the integration proceeds from blue to red. The bottom panel of Fig.~\ref{fig:noise} illustrates the $\overline{B}^{\,\prime}_z$ CSV observable normalised by the residual profile intensity $1-{\cal R}_I$ according to Eq.~(\ref{eq19}). The devastating impact of the noise at profile edges is apparent.

It is possible to define the CSV profiles using an alternative, backward, red-to-blue integration scheme
\begin{equation}
{\cal G}_V^{-} (\lambda) = - \dfrac{1}{W_I} \int^{\lambda_{\rm max}}_{\lambda} {\cal R}_V(\lambda^\prime) d \lambda^\prime.
\label{eq22}
\end{equation}
This formula gives identical results to Eq.~(\ref{eq14}) in the absence of noise. When the noise is present, it will lead to a systematic discrepancy similar to the one illustrated in Fig.~\ref{fig:noise}, but increasing towards the blue side of the line. The overall systematic deviation from zero can be minimised by a weighted mean of the forward and backward integrations
\begin{equation}
\overline{{\cal G}_V} (\lambda) = \dfrac{1}{W_I} \left[ 
 w^{+} \int_{\lambda_{\rm min}}^{\lambda} {\cal R}_V(\lambda^\prime) d \lambda^\prime
- w^{-} \int^{\lambda_{\rm max}}_{\lambda} {\cal R}_V(\lambda^\prime) d \lambda^\prime \right],
\label{eq24}
\end{equation}
where the weights $w^{\pm}$ are given by
\begin{equation}
w^{+} = \dfrac{\lambda_{\rm max} -\lambda}{\lambda_{\rm max} - \lambda_{\rm max}}
\mathrm{\quad and \quad}
w^{-} = \dfrac{\lambda-\lambda_{\rm min}}{\lambda_{\rm max} - \lambda_{\rm max}}.
\label{eq24a}
\end{equation}
The correction expressed by these equations is equivalent to subtracting from the original CSV profile ${\cal G}_V (\lambda)$ a straight line going through 0 at $\lambda_{\rm min}$ and ${\cal G}_V (\lambda_{\rm max})$  at $\lambda_{\rm max}$.

Application of Eqs.~(\ref{eq24}--\ref{eq24a}) still results in a substantially higher reduced $\chi^2$ than found for the original Stokes $V$ profiles corrupted by noise. Using $10^6$ realisations of normally-distributed random noise in the absence of any background signal, we have built a cumulative distribution function for the $\chi^2$ values deduced by performing integration according to Eq.~(\ref{eq24}) with associated error propagation. As shown in the upper panel of Fig.~\ref{fig:chi}, integration leads to substantially larger $\chi^2$ values compared to the initial Stokes $V$ profiles. For example, the $10^{-3}$ probability threshold corresponds to $\chi^2_\nu\approx6$ for the CSV profiles compared to $\chi^2_\nu\approx1.4$ for the initial profiles.

Based on this empirical CDF of $\chi^2_\nu$ of CSV profiles, we estimated the fraction of such profiles that will be detected above the noise level with the false alarm probability of $10^{-3}$ or smaller for the four-spot magnetic field geometry sampled at 100 equidistant rotational phases. These MC simulations were carried out for noise amplitudes in the range from $2\times10^{-4}$ to $2\times10^{-3}$, i.e. 10--100\% of the mean peak-to-peak Stokes $V$ profile amplitude. Simulations were repeated $10^4$ times for each value of the noise amplitude.

As can be seen from the lower panel of Fig.~\ref{fig:chi}, the detection of magnetic field signatures using CSV profiles is less successful than with the original Stokes $V$ spectra. Therefore, the CSV diagnostic technique does not represent a useful alternative, in terms of the mere field detection, to the widely used $\chi^2$-based Stokes $V$ profile detection method.

\section{Observational data}
\label{obs}

For the purpose of testing the CSV diagnostic method on real data we analysed high-resolution spectropolarimetric observations available from the {\sc Polarbase} archive \citep{petit:2014}. This database provides reduced, calibrated, one-dimensional Stokes parameter spectra obtained with ESPaDOnS at CFHT and Narval at TBL of the Pic du Midi observatory. Both instruments are fibre-fed, thermally stabilised, echelle spectropolarimeters covering the wavelength range from 3694 to 10483~\AA\ in a single exposure at the resolving power of $\lambda/\Delta\lambda=65\,000$.

As an example of the Stokes $V$ profiles of a cool active star with a complex magnetic field we considered spectropolarimetric observations of the weak-line T~Tauri star \vtau. This data set, discussed in the papers by \citet{skelly:2010} and \citet{rice:2011}, comprises 45 Stokes $V$ observations obtained in the span of about two weeks in January 2009. The signal-to-noise ratio of these spectra, S/N\,$\approx$\,150, is insufficient for the detection and analysis of circular polarisation signatures in individual spectral lines. Therefore, we applied the least-squares deconvolution approach \citep{kochukhov:2010a} to each spectrum, combining about 4400 spectral lines into a single high S/N ratio profile. The data were phased with the ephemeris of \citet{stelzer:2003}.

We also analysed the Stokes $V$ observations of the He-strong Bp star \hd. This object is known to possess a complex magnetic field, strongly deviating from an oblique dipolar geometry \citep{kochukhov:2011a}. There are 27 Stokes $V$ spectra of \hd, with a typical S/N ratio of 450, available in {\sc Polarbase}. These data were acquired over the period of 2006--2012. We used the variable-period ephemeris of \citet{mikulasek:2008} to compute rotational phases. Since the magnetic field of \hd\ is very strong, circular polarisation signatures are readily observable in individual spectral lines. In this paper we study the \ion{He}{i} 6678~\AA\ line. Its circular polarisation shows a variability pattern very similar to the \ion{He}{i} 5876~\AA\ line analysed by \citet{kochukhov:2011a} using a lower quality data.

\begin{figure*}[!th]
\centering
\finps{!}{16cm}{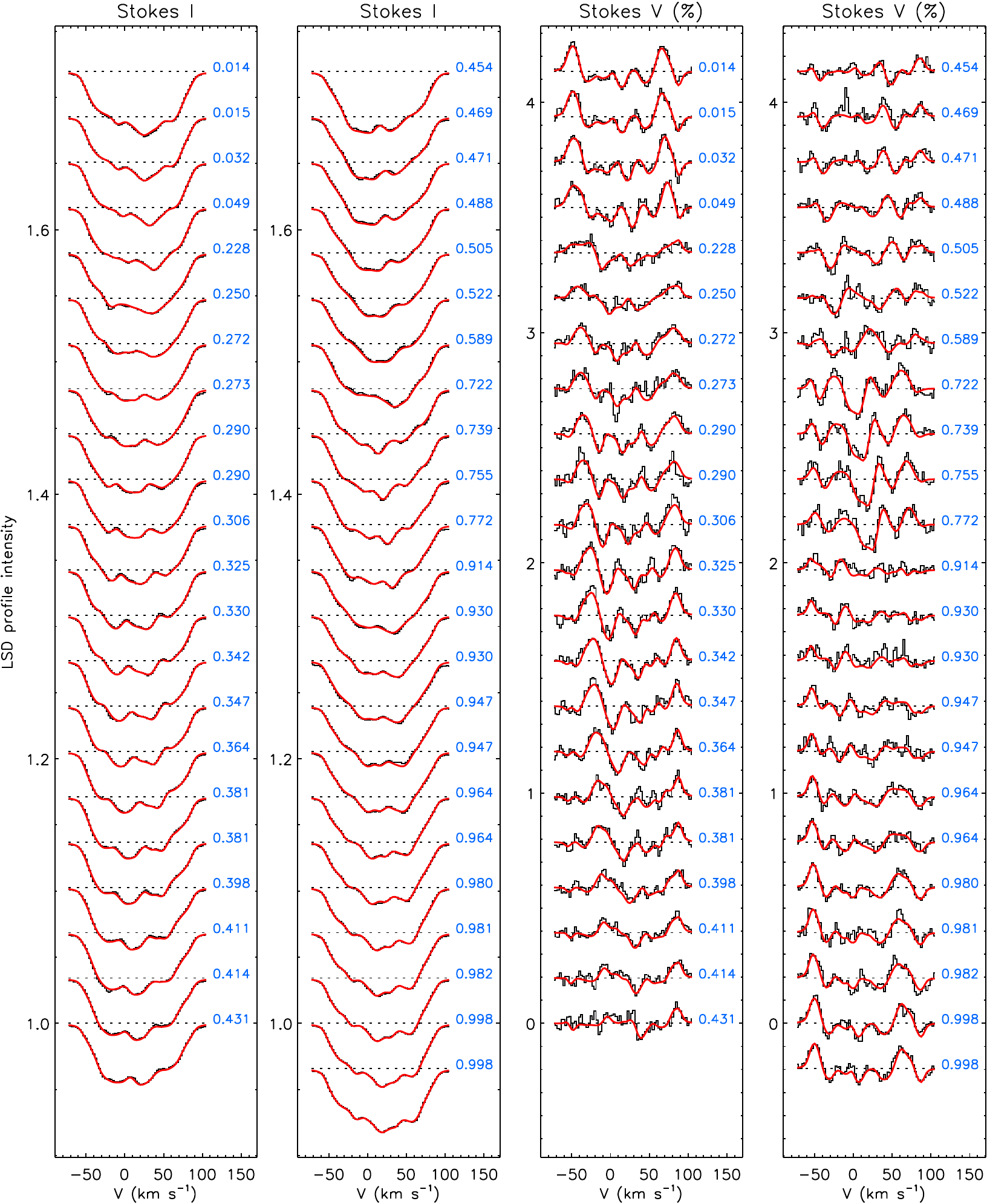}\hspace*{0.5cm}
\finps{!}{16cm}{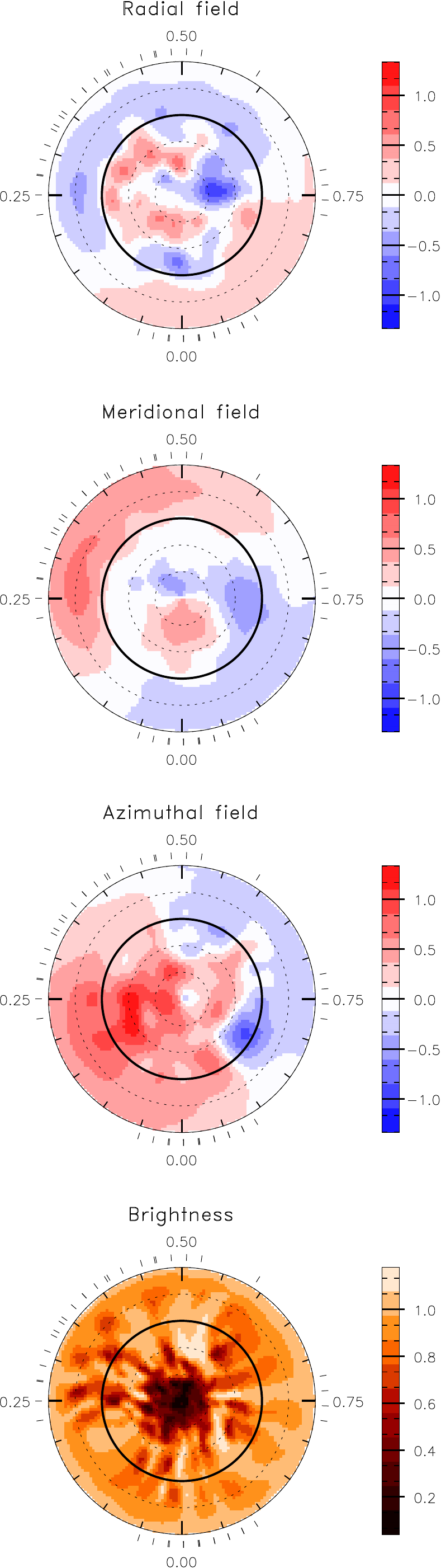}
\caption{Results of the ZDI analysis of \vtau. The four panels on the left side of the figure compare the observed (histogram) and theoretical (solid line) LSD Stokes $I$ and $V$ profiles. In these plots the spectra corresponding to different rotational phases are shifted vertically. The phase is indicated to the right of each profile. Reconstructed maps of the radial, meridional and azimuthal magnetic field components as well as the brightness distribution are presented in the right column. The star is shown using the flattened polar projection between latitudes $-60\degr$ and $+90\degr$. The thick circle corresponds to the rotational equator. The colour bars give the field strength in kG and the brightness relative to the photospheric value.}
\label{fig:ZDI}
\end{figure*}

\section{Examples of the CSV diagnostic}

\subsection{Weak-line T~Tau star \vtau}

\begin{figure*}[!th]
\centering
\firps{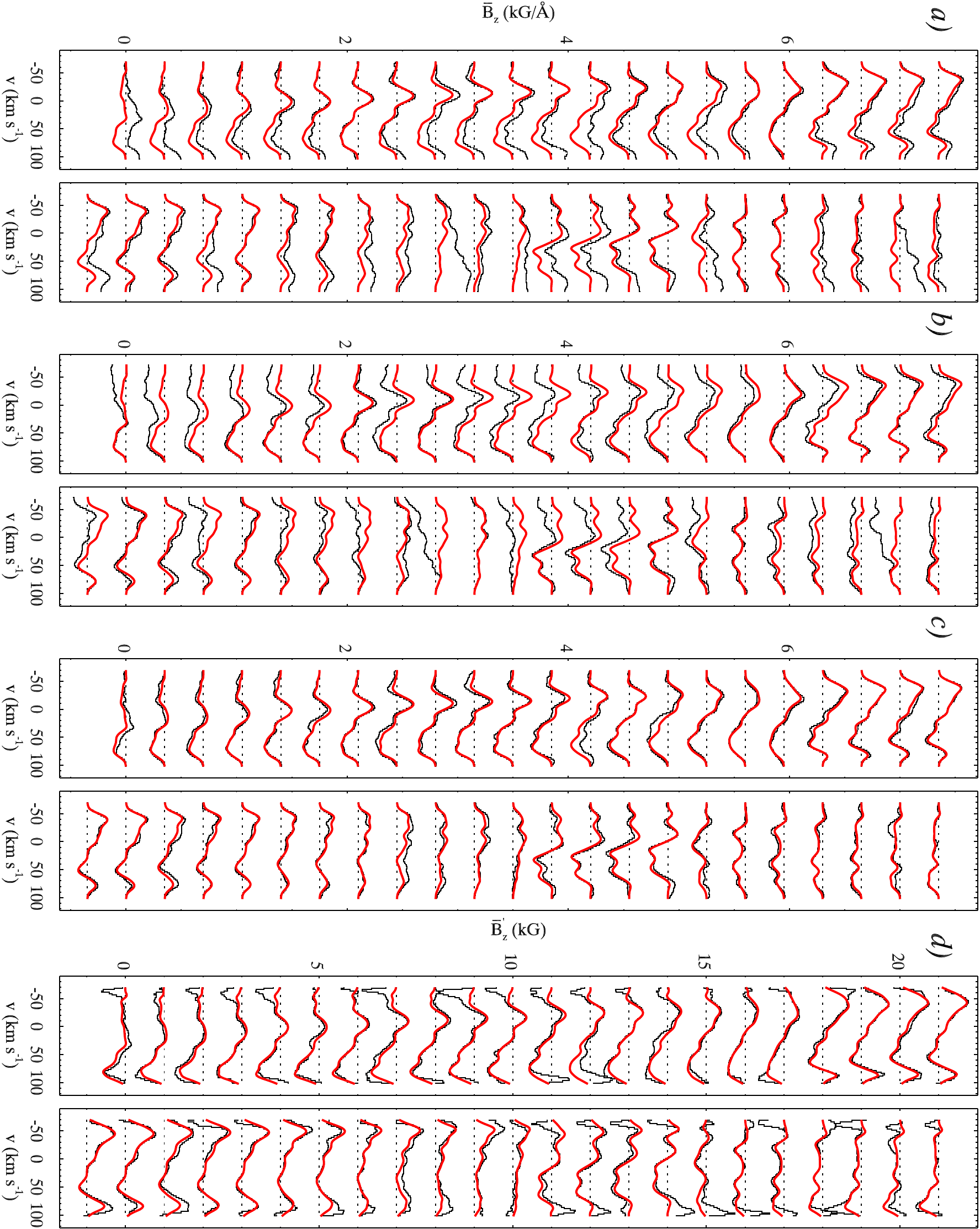}
\caption{Cumulative Stokes $V$ profiles of \vtau\ computed from the observed (histogram) and theoretical (thick solid line) LSD circular polarisation spectra of \vtau\ shown in Fig.~\ref{fig:ZDI}. The CSV profiles are shifted vertically according to their rotational phases similar to the Stokes $V$ profiles in Fig.~\ref{fig:ZDI}. The first three panels show the CSV spectra computed with {\bf a)} forward integration, {\bf b)} backward integration, and {\bf c)} weighted mean of the backward and forward integration. The last panel {\bf d)} shows the result of converting the CSV profiles in panel {\bf c)} to the $1-{\cal R}_I$ normalisation according to Eq.~(\ref{eq19}).}
\label{fig:observed_CSV}
\end{figure*}

\begin{figure*}[!th]
\centering
\finps{!}{9cm}{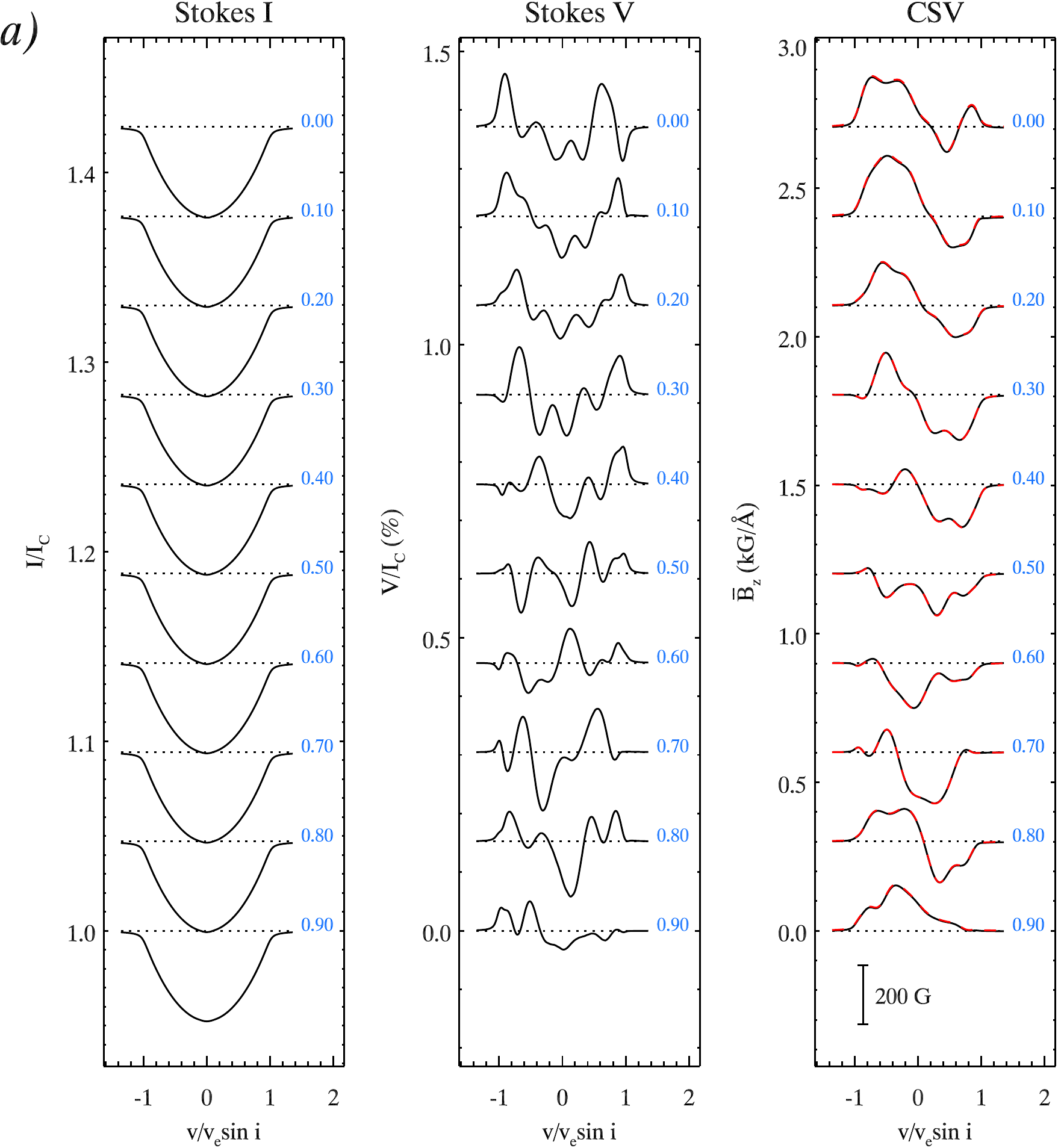}\hspace*{0.5cm}
\finps{!}{9cm}{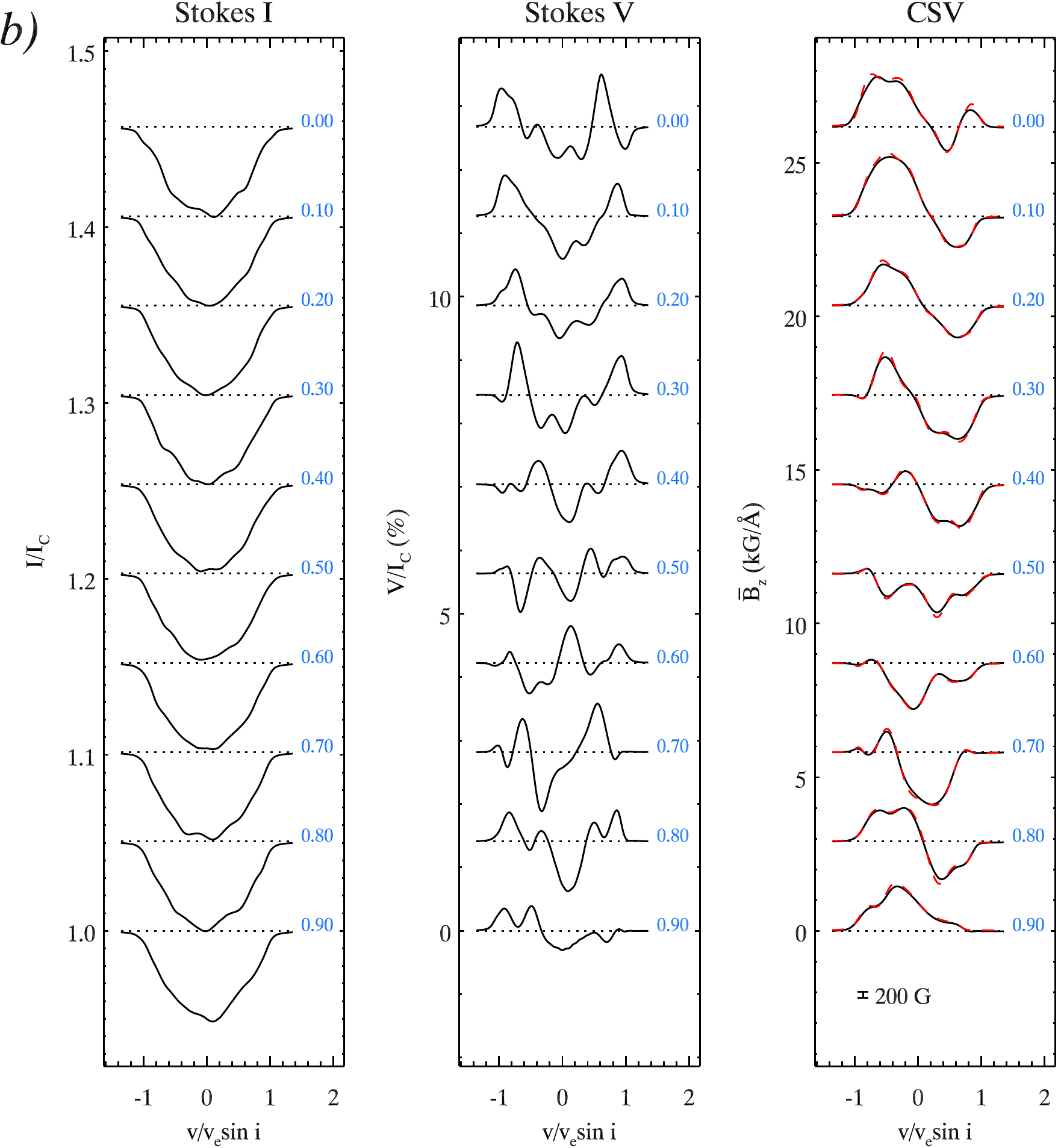}
\caption{Theoretical Stokes $I$, $V$ and CSV profiles corresponding to {\bf a)} the magnetic field map of \vtau\ shown in Fig.~\ref{fig:ZDI} and {\bf b)} the same map scaled by a factor of 10. In each case the CSV column compares $\overline{B}_z(\lambda)$ obtained from the Stokes $I$ and $V$ line profiles (solid line) with the results of numerical integration of the line-of-sight magnetic field component over the stellar disk (dashed line).}
\label{fig:scaled_CSV}
\end{figure*}

The young rapidly rotating star \vtau\ (HD\,283518) exhibits ample signs of the surface magnetic activity and was targeted by a number of temperature DI studies \citep{hatzes:1995,rice:1996,rice:2011}. These analyses revealed a persistent large polar spot and evolving temperature inhomogeneities at lower latitudes. Magnetic field was detected on \vtau\ by \citet{donati:1997}. \citet{skelly:2010} reconstructed the surface magnetic field topology of this star with ZDI, finding a significant toroidal magnetic component and a complex radial field distribution. On the other hand, \citet{carroll:2012} reported a relatively simple, predominantly poloidal magnetic field structure consisting of two large polar magnetic spots with an opposite polarity. The Stokes $V$ spectropolarimetric observations interpreted by \citet{skelly:2010} and \citet{carroll:2012} were obtained with the twin instruments (ESPaDOnS and Narval) and at practically the same time, meaning that the discrepant magnetic inversion results cannot be ascribed to the intrinsic variation of the surface magnetic field of \vtau.

Here we use the combined ESPaDOnS and Narval spectropolarimetric data set to reconstruct independent maps of the magnetic field topology and brightness distribution. The inversion methodology applied to \vtau\ is described in detail by \citet{kochukhov:2014}. Briefly, the magnetic field of the star is represented in terms of a superposition of the poloidal and toroidal harmonic terms with the angular degree up to $\ell=15$. A penalty function prohibits unnecessary contribution of the high-order harmonic modes. A separate regularisation procedure applied to the brightness map minimises any deviation from the reference (photospheric) brightness value. We employed the analytical Unno-Rachkovsky Stokes parameter profiles \citep[e.g.][]{polarization:2004} to approximate the local intensity and circular polarisation spectra. The brightness and magnetic field maps were recovered self-consistently from the LSD Stokes $I$ and $V$ profiles, adopting $i=60\degr$ and $v_{\rm e}\sin{i}=74$~\kms.

The LSD profile fits and the resulting brightness and magnetic field maps are presented in Fig.~\ref{fig:ZDI}. The surface distributions which we obtain are generally compatible with the maps published by \citet{skelly:2010}. Similar to these authors we find a dominant dark polar spot and numerous small-scale brightness inhomogeneities at lower latitudes. The poloidal and toroidal field components contribute 63\% and 37\% respectively to the total magnetic field energy. The field strength reaches $\sim$\,1~kG locally on the stellar surface. The azimuthal field exhibits two large unipolar regions while the radial field shows a more complex structure. The darkest areas in the brightness map do not coincide with any particular magnetic field features. We have also verified that a very similar magnetic field geometry is obtained if one performs a direct ZDI reconstruction of each magnetic field component without using the spherical harmonic formalism.

Based on the magnetic inversion results, we calculated the CSV profiles using both the observed and the ZDI model Stokes $I$ and $V$ spectra of \vtau. Figure~\ref{fig:observed_CSV} compares profiles obtained by applying Eq.~(\ref{eq14}), Eq.~(\ref{eq22}), and the weighted mean of the forward and backward integrations given by Eq.~(\ref{eq24}). It is clear that, in the first and second case (Figs.~\ref{fig:observed_CSV}ab), a significant systematic deviation between the observed and theoretical CSV profiles appears due to the noise accumulation. On the other hand, systematic effects are largely brought under control in the third set of CSV profiles (Fig.~\ref{fig:observed_CSV}c).

Figure~\ref{fig:observed_CSV}d illustrates the impact of employing the $1-{\cal R}_I$ normalisation in place of the $W_I$ normalisation used elsewhere. The $\overline{B}^{\,\prime}_z$ profiles shown in this panel directly provide the average line-of-sight magnetic field component at different longitudes on the stellar surface. However, these profiles, being morphologically similar to the $\overline{B}_z$ spectra, are visibly distorted by the noise artefacts at the line edges.

We tested the accuracy of the CSV diagnostic with the model LSD profiles corresponding to the magnetic field map shown in Fig.~\ref{fig:ZDI}. For clarity we disregarded the non-uniform brightness distribution and considered the line profiles for 10 equidistant rotational phases. The CSV profiles were computed from the simulated observations using Eq.~(\ref{eq14}) and by performing a numerical integration of the line-of-sight magnetic field component according to the right-hand side of Eq.~(\ref{eq17}). In these calculations we used the same local non-magnetic $r_I$ profile as was adopted for the ZDI inversions. As expected, the two sets of CSV profiles agree perfectly (Fig.~\ref{fig:scaled_CSV}a).

The CSV method is based on the weak-field approximation, which is usually considered to be valid only for the magnetic field strengths below $\sim$\,1~kG. It is useful to assess the usefulness of the cumulative Stokes $V$ profiles for much stronger magnetic fields. To this end, we scaled all the vector components of the ZDI map of \vtau\ by a factor of 10. This increased the mean field strength from 570~G to 5.7~kG. We then repeated the direct and geometrical evaluation of the CSV profiles. The resulting spectra are compared in Fig.~\ref{fig:scaled_CSV}b. The two sets of $\overline{B}_z$ profiles show only marginal discrepancies despite that the characteristic magnetic field strength now significantly exceeds 1~kG. Therefore, it appears that the CSV diagnostic is usable well beyond the nominal limit of the weak-field approximation.

\begin{figure}[!th]
\centering
\fifps{\hsize}{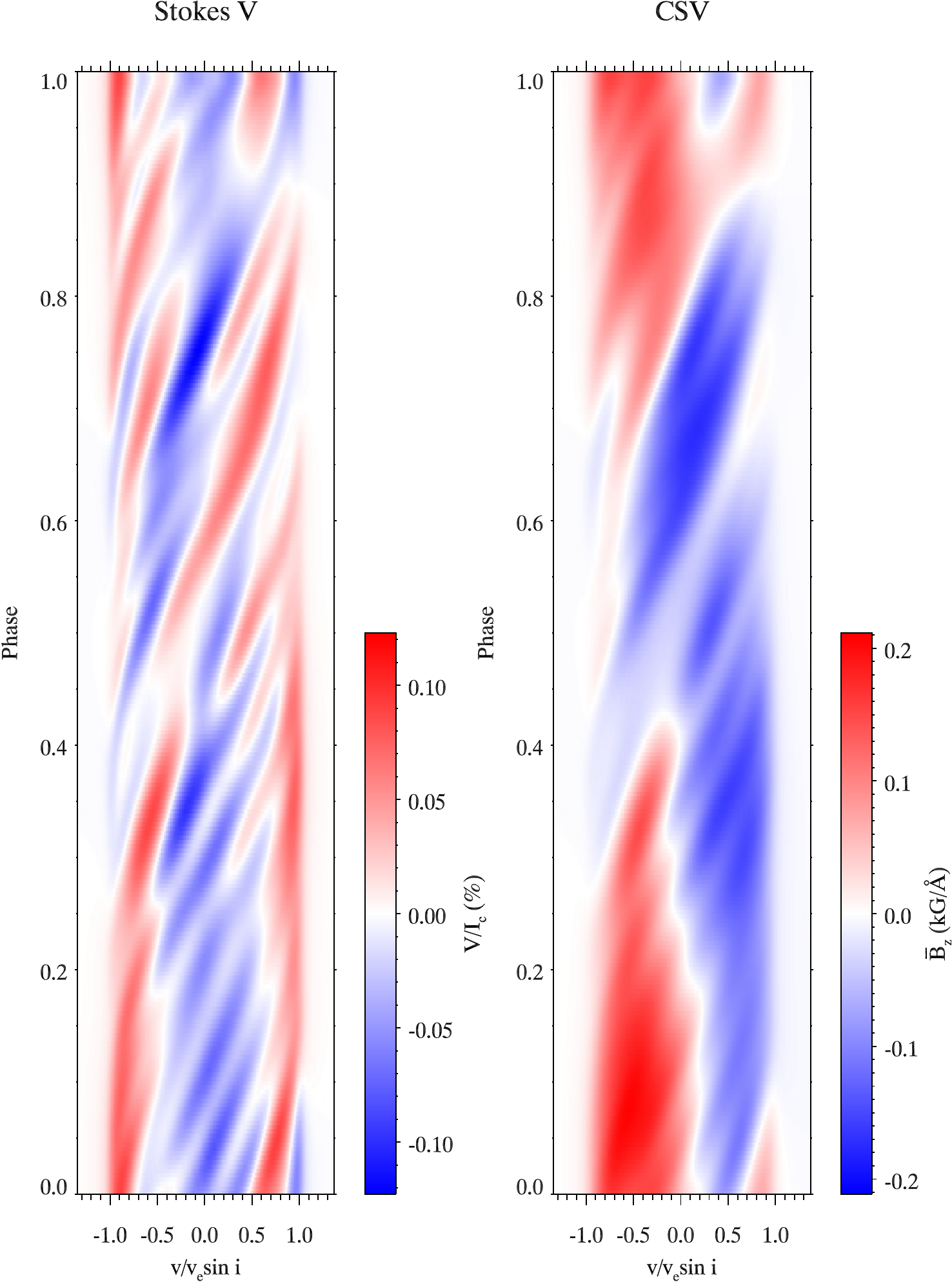}
\caption{Theoretical dynamic Stokes $V$ (left panel) and cumulative Stokes $V$ (right panel) profiles corresponding to the ZDI map of \vtau\ shown in Fig.~\ref{fig:ZDI}.}
\label{fig:dynamic_v410}
\end{figure}

\begin{figure}[!th]
\centering
\fifps{\hsize}{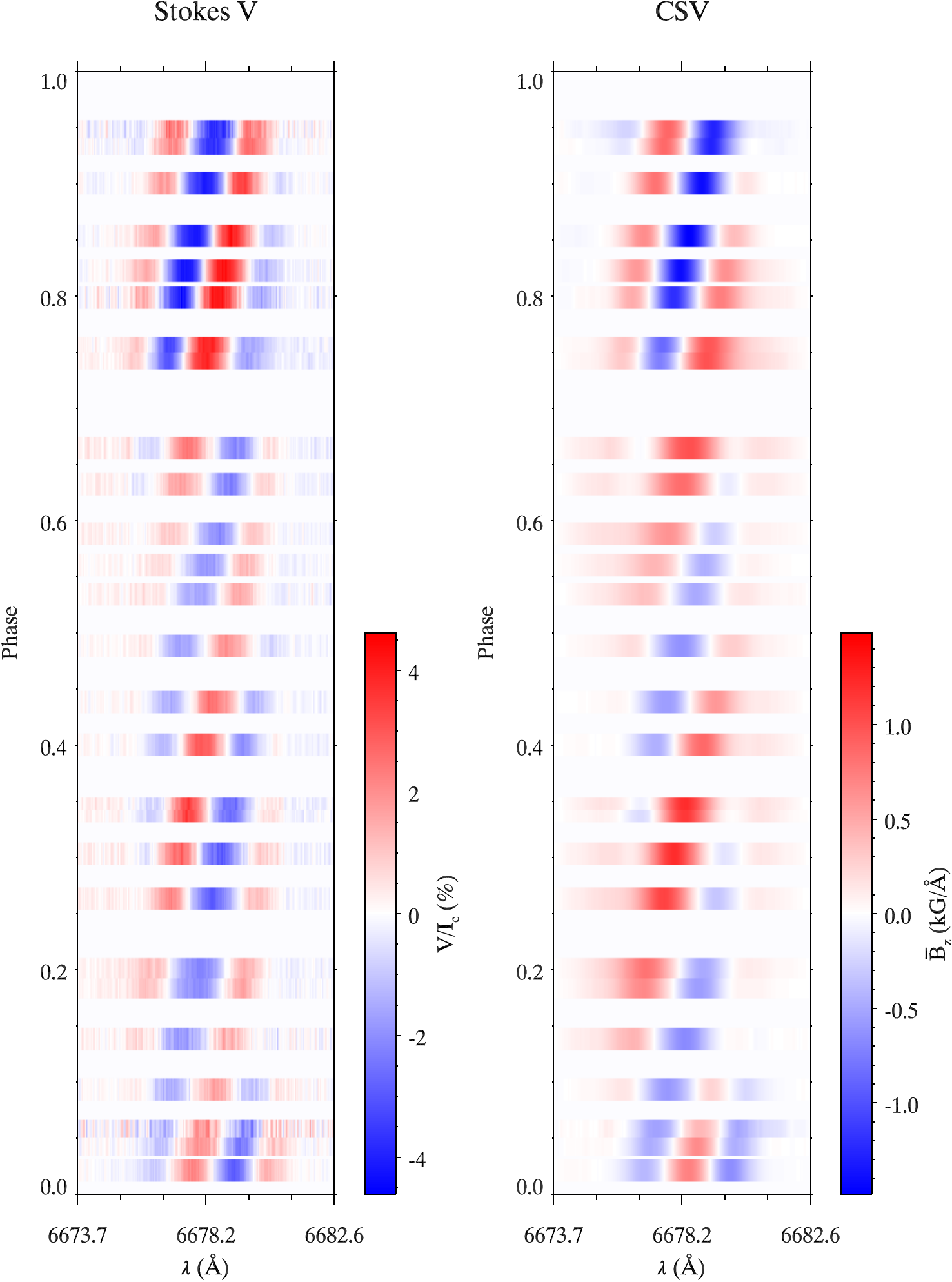}
\caption{Observed dynamic Stokes $V$ (left panel) and cumulative Stokes $V$ (right panel) profiles of the \ion{He}{i} 6678~\AA\ line in Bp star HD\,37776.}
\label{fig:dynamic_ap}
\end{figure}

Using the ZDI maps of \vtau\ we calculated variation of the Stokes parameter profiles with a dense rotational phase sampling. The resulting dynamic circular polarisation and CSV spectra are presented in Fig.~\ref{fig:dynamic_v410}. Unlike Fig.~\ref{fig:dynamic_model} it shows a stationary pattern (phases 0.0--0.3) in addition to features travelling across the stellar disk (e.g. phase 0.6--0.8). The travelling features are associated with the radial field spots, in particular the negative polarity spot best visible at the rotational phase 0.75. The stationary features correspond to the large unipolar azimuthal fields regions. Such behaviour of the dynamic CSV profiles can be considered as an evidence for the presence of a significant toroidal magnetic field on the stellar surface.

\subsection{He-strong Bp star HD\,37776}

The B-type magnetic chemically peculiar star \hd\ (V901 Ori) is known for its unusually complex, double-wave longitudinal magnetic field variation \citep{thompson:1985}. This behaviour indicates a large deviation of the stellar surface magnetic field topology from an oblique dipolar geometry. In the literature this object is often considered to be an example of the star with a quadrupolar magnetic field structure. Indeed, \citet{bohlender:1994} and \citet{khokhlova:2000} derived quadrupolar magnetic geometry models for \hd\ based on fitting the longitudinal field curve and Stokes $V$ profile variation, respectively. However, using a more extensive set of phase-resolved spectropolarimetric observations, \citet{kochukhov:2011a} showed that the circular polarisation spectra of \hd\ cannot be fitted with an axisymmetric quadrupolar field structure. Instead, the ZDI inversion carried out in that study suggested a complex field distribution comprising a series of magnetic spots with alternating positive and negative polarity.

The high-resolution circular polarisation spectra of \hd\ analysed here enable a straightforward verification of the ZDI results of \citet{kochukhov:2011a} using the cumulative Stokes $V$ profile diagnostic method. Figure~\ref{fig:dynamic_ap} shows the Stokes $V$ and CSV spectra centred on the \ion{He}{i} 6678~\AA\ line. The dynamic CSV profiles are dominated by a series of travelling features. There is no evidence of stationary structures like in the case of \vtau, indicating that no strong toroidal magnetic field is present. Furthermore, a careful examination of the CSV profiles allows one to identify three spots with negative polarity and the same number of positive-field features. This corresponds very well to the six well-defined spots seen in the radial magnetic field map recovered with ZDI \citep[see fig.~5][]{kochukhov:2011a}. On the other hand, considering the dynamic CSV profiles in Fig.~\ref{fig:dynamic_ap} one can immediately rule out the quadrupolar field hypothesis. A quadrupolar magnetic field geometry with a large obliquity, as required to reproduce the \bz\ variation of \hd, exhibits four regions of alternating field polarity along the stellar rotational equator. But Fig.~\ref{fig:dynamic_ap} shows at least six regions.

\section{Summary and discussion}

In this paper we examined the information content of the cumulative circular polarisation profiles and assessed their usefulness for the analysis of stellar surface magnetic fields. This work represents a further development of the idea put forward by \citet{gayley:2014}. Starting with the weak-field approximation, we gave a definition of the new magnetic observable as a normalised integral of the Stokes $V$ spectrum and showed its relation to the underlying surface magnetic field structure. The transformation from the Stokes $V$ to the CSV profiles greatly simplifies the morphology of polarisation spectra and enables a direct and intuitive identification of the regions of different field polarity on the stellar surface. The CSV observable characterises the line-of-sight magnetic field component weighted by the continuum intensity and by the Doppler-shifted local residual Stokes $I$ profile. The CSV spectrum thus provides a velocity-resolved measure of the stellar longitudinal magnetic field. Consequently, this observable is most useful for rapidly rotating stars with complex magnetic field topologies.

Using theoretical circular polarisation profile calculations and real observational data we assessed visibility of the surface magnetic field features in the dynamic CSV profiles. Unlike the dynamic Stokes $V$ spectra, which can hardly be interpreted directly, the dynamic CSV profiles enable a qualitative analysis of the magnetic star spot distributions. With the help of Monte Carlo simulations we studied the effect of a random observational noise on the CSV profiles. The noise signatures in the CSV spectra are highly correlated, leading to a systematic ramping offset from a noise-free version of the line profiles. We showed that this problem can be minimised using a weighted mean of the CSV profiles computed with the forward and backward integration of the original Stokes $V$ data. Nevertheless, the presence of correlated noise in the CSV profiles generally makes them inferior for the purpose of field detection compared to the original Stokes $V$ spectra. 

With the goal of testing the CSV magnetic field diagnostic methodology we studied the archival high-resolution circular polarisation observations of the cool active star \vtau\ and the hot magnetic star \hd. Both objects are rapid rotators with complex surface magnetic field topologies. For the weak-line T~Tauri star \vtau\ we obtained brightness and magnetic field distributions with the help of the ZDI modelling of the least-squares deconvolved Stokes $I$ and $V$ profiles. The resulting magnetic map bears a close resemblance to the magnetic field topology obtained by \citet{skelly:2010} from the same data. However, our inversion results contradict the magnetic field map published by \citet{carroll:2012} whose data set was also included in our analysis. Our ZDI modelling accounted for the effect of an inhomogeneous brightness distribution on the Stokes $V$ profiles. We have also verified that very similar magnetic field maps are obtained using the direct, pixel-based ZDI and the magnetic inversions relying on the spherical harmonic parameterisation of the stellar magnetic field. Therefore, these two aspects are unlikely to be responsible for the discrepancies of the ZDI maps presented here and in \citet{skelly:2010} on the one hand and in \citet{carroll:2012} on the other hand. It is more likely that this disagreement is rooted in the differences between the LSD \citep[used here and by][]{skelly:2010} and PCA \citep[used by][]{carroll:2012} mean polarisation profiles. A comprehensive comparative study of the LSD and PCA line-addition techniques is required to address this problem.

The dynamic CSV profiles of \vtau\ indicate the presence of a strong azimuthal magnetic field component on the stellar surface. This agrees with our ZDI results and suggests a non-negligible contribution of the toroidal magnetic field. We also used the ZDI map of \vtau\ to verify interpretation of the integral of Stokes $V$ profiles in terms of the disk-averaged, weighted line-of-sight magnetic field component. These calculations were repeated for a scaled-up version of the stellar magnetic field geometry, demonstrating that the usefulness of the CSV diagnostic extends far beyond the nominal limit of the weak field approximation.

For the magnetic Bp star \hd\ we analysed the CSV dynamic spectrum of the \ion{He}{i} 6678~\AA\ line. The variation of its CSV profiles is clearly inconsistent with the quadrupolar magnetic field topology frequently suggested for this star. On the other hand, multiple magnetic spots which can be identified in the dynamic CSV spectra of \hd\ are consistent with the surface magnetic features in the ZDI map of this star derived by \citet{kochukhov:2011a}. The CSV analysis thereby confirms the complex, non-quadrupolar nature of the magnetic field topology of \hd.

\begin{acknowledgements}
This research is supported by the grants from the Knut and Alice Wallenberg Foundation, the Swedish Research Council, and the Swedish National Space Board. The author thanks Dr. K. Gayley for helpful discussions of the CSV diagnostic method and Dr. J. Silvester for critical reading of manuscript.
\end{acknowledgements}

%\bibliographystyle{aa}
%\bibliography{astro_papers}

\end{document}